# Negative Refraction in Perspective

A D BOARDMAN, N KING, L VELASCO,
Joule Physics Laboratory
Institute for Materials Research
University of Salford
Salford, M5 4WT, UK

*The concept of negative refraction is attracting a lot of attention. The initial ideas and the misconceptions that have arisen are discussed in sufficient detail to understand the conceptual structure that binds negative refraction to the existence of backward wave and forward wave phenomena. A presentation of the properties of isotropic media supporting backward waves is followed by discussion of negative phase media, causality, anisotropic crystals and some connections to photonic crystals. The historical background is always coupled to a detailed presentation of all the issues. The paper is driven numerically and is lavishly illustrated with the outcomes of original FDTD simulations.*

## Introduction

There is a rapidly growing literature that uses the word *negative* (Pendry 2003,2004; Lahthakia 2003a, 2003b; McCall, Lahthakia & Weiglhofer 2002; Lindell et al. 2001; Ziolkowski & Kipple 2003; Smith, Schurig & Pendry 2002; Boardman et al. 2005;) in connection with phase velocity, or refraction.  In addition, this adjective is quite often associated with the materials that are called *left-handed* and the refraction they display.   Many articles appear to discuss the topic of this paper in the correct kind of way but some go on to misdirect by inappropriate associations embedded in what is often an inappropriate terminology.   On the face of it, it is a daunting task to give a satisfactory perspective on the topic of negative refraction so perhaps it is wise to take a leaf out of Lewis Carroll's book "Alice's Adventures in Wonderland" and simply take the view that it is best to begin at the beginning, go on until the end is reached and then stop. The temptations to engage in elegant but abstract descriptions of a theoretical kind, or just encyclopaedic descriptions, will be avoided in favour of a selective offering that makes important points. The arguments are visually supported, sometimes at a very advanced computational level.

The beginning belongs to the great optical scientist Sir Arthur Schuster who, as Professor of Physics in the University of Manchester, in 1904, was a colleague of the famous hydrodynamicist Sir Horace

Lamb. It is not possible to be absolutely categorical about the conversations that they had but the outcome that matters to us is crystal clear. Lamb was then concerned about the relationship between the group velocity (Lamb 1916) and individual, or phase, velocity of the waves bundled together as a group. First of all, it is very easy to witness group and phase velocity in action, simply by throwing a pebble into a pond, or lake, and then watching the ripples spread out. The disturbance appears to be limited in space and time and is in fact a bunch, or group, of faster or slower waves that then travel through it. In practice, these individual waves appear at one end of the group, travel across it and then simply die out at the other end. It is intuitively obvious that energy is carried by the group and that this energy is being transported at the group velocity. Lamb also recognised that the individual waves in a group can, in principle, also set out at the front of the group and die out at the rear of it. Schuster then considered (Schuster 1904) this type of *backward wave* phenomenon in an optical context. He pointed out that backward waves may be generated whenever light propagates within an optical absorption band. In such regions, what came to be known, inappropriately, as anomalous dispersion occurs. All materials have one or more bands like this, however, so there is nothing unusual about materials like fuchsin and cyanin (figure 1) that were nominated by Schuster as examples of this phenomenon. In an absorption region the wave velocity increases as the wavelength decreases and the group velocity can be anti-parallel to the phase (individual) velocity. However, Schuster did not discuss the kind of possibilities outlined below.

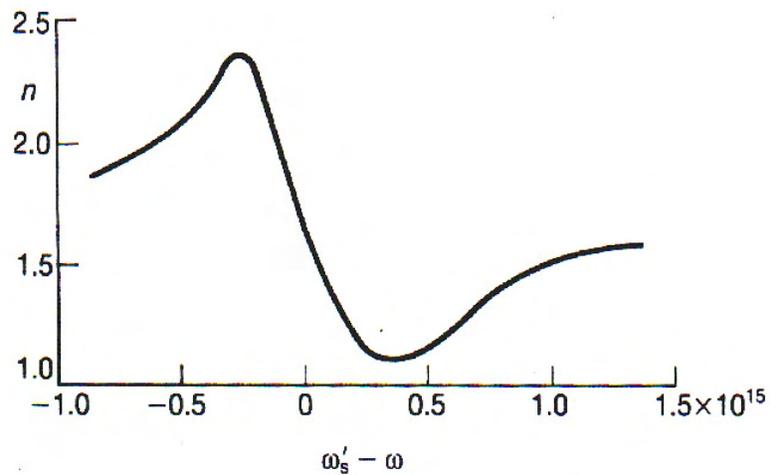

Figure 1: Dispersion curve for cyanin showing a region of anomalous dispersion. Note that the curve is displayed by taking the origin at a frequency $\omega_s'$

Schuster found this all very fascinating but was unimpressed by the fact that massive absorption implies that material thicknesses of less than a wavelength would be forced upon any experimental test for backward waves. Apart from this caveat, Schuster produced an impressive diagram (Figure 2) showing what will happen when a plane wave of light is incident upon an interface of a semi-infinite backward wave medium. Schuster states that "One curious result follows: the deviation of the wave entering such a medium is greater than the angle of incidence".

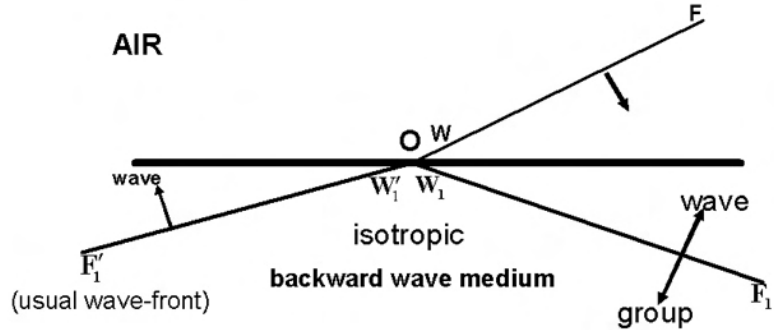

Figure 2: A redrawn, edited, form of Schuster's original diagram in which he shows the incident wave front WF. He also shows the direction of propagation, in the usual way, by the arrow that is perpendicular to a wave front.

There are two possibilities for the refracted wave front i.e. $W_1F_1$ or $W_1'F_1'$. Schuster noted that the problem with using $W_1'F_1'$ is that its intersection with the surface slides to the left while the intersection of the incident wave front slides to the right. This is because the material sustains phase waves that are propagating towards the surface. In order to lock these intersection points together it is necessary to adopt the refraction wave front $W_1F_1$. In this way, Schuster was able to conclude that energy can be carried forward at the group velocity but in a direction that is anti-parallel to the phase velocity. Given this conclusion, an incident light wave looks as though it refracts past the normal and this aroused Schuster's curiosity. It is interesting that this conclusion was reached for a purely dielectric material, characterised by relative dielectric permittivity and a relative permeability equal to that of free space. In other words, he saw that there was no necessity to invoke a dispersive relative magnetic permeability, as is the case for the so-called left-handed media to be addressed below.

Energy is transported at the speed of the group and the convention that it is being transported in a "*positive*" direction can be taken. It is then logical to assert that in the *isotropic* medium discussed by Schuster that the phase velocity is travelling in a "*negative*" direction. Of course, there is a kind of upbeat arbitrariness to this convention because all that is being conveyed is that in an isotropic backward wave medium the energy flow and the phase velocity direction are anti-parallel. Nevertheless, the advocacy that this type of isotropic medium should be referred to as a *negative phase velocity* medium carries considerable conviction. A second point concerns the refraction itself. Again it is not unreasonable to refer to the normal, everyday, refraction that everybody has witnessed as a manifestation of *positive refraction* so that any refraction that goes beyond the normal could be called *negative refraction*. It does not follow that it is necessary to describe this behaviour in terms of an actual negative refractive index, but this issue will be returned to later on in the paper. In the Schuster example, negative refraction is clearly linked to the negative phase velocity, discussed earlier. However, much of the modern literature has taken the two things to be entirely synonymous. This is not always the case and it will be seen later that there are very common examples of negative refraction, in which the phase wave and group travel in a *forward*, but not necessarily the same, direction. A note of caution then is to avoid always associating negative refraction with the presence of backward waves.

Returning briefly to absorption bands, Figure 3 shows that a negative gradient in the refractive index variation with frequency does not necessarily imply that the group and phase velocity are oppositely directed. The relationship between a dispersion curve $(\omega, k)$, which relates angular frequency $\omega$ to the wave number *k* and the frequency dependence of the index of refraction, namely $(n, \omega)$, is simply $n = \dfrac{ck}{\omega}$ so that

$$\frac{d\omega}{dk} = \frac{c}{\omega}\left[\frac{n}{\omega} + \frac{dn}{d\omega}\right]^{-1} \tag{1}$$

and

$$\frac{dn}{d\omega} < 0 \quad \text{and} \quad \left|\frac{dn}{d\omega}\right| > \frac{n}{\omega} \tag{2}$$

These inequalities show that, if a negative group velocity is required, the gradient of the $(n,\omega)$ curve must not only be negative but the magnitude of the gradient of the $(n,\omega)$ curve must also be greater than $n/\omega$. If this is established for a given absorption curve then it will define the region in which the group and phase velocity are anti-parallel and the sign convention of positive energy flow can be adopted as explained earlier. The fundamental point is that backward waves can be supported and hence negative refraction can be expected. It is important to recognise from this discussion that a negative gradient in $(n,\omega)$ curve could also be associated with a positive gradient in the $(\omega,k)$ curve coupled to the fact that the group velocity is greater than the phase velocity. This case is most often referred to as anomalous dispersion. The cases just discussed are illustrated in Figure 3.

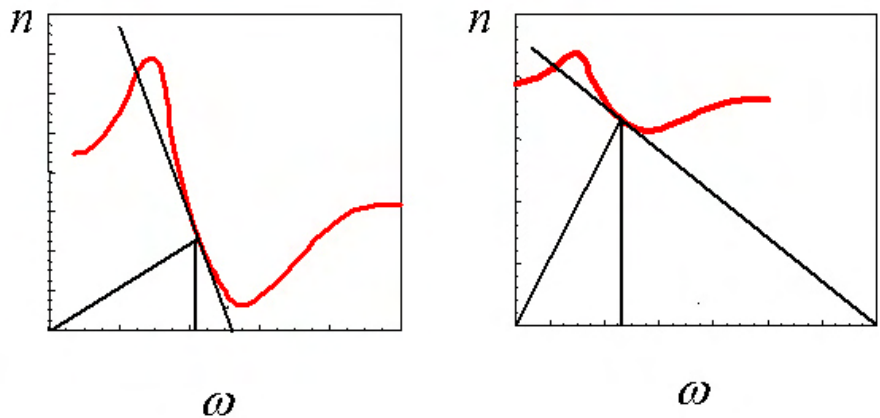

Figure 3: Sketches of the refractive index n versus frequency $\omega$. (a) magnitude of gradient of $dn/d\omega$ *greater* than $n/\omega$. Anomalous dispersion with *positive* phase velocity and *negative* group velocity but $|v_g| > v_p$ b) Gradient of $dn/d\omega$ *less* than $n/\omega$. Anomalous dispersion with *positive* phase velocity and *positive* group velocity and $v_g > v_p$

Now that some of the background has been established this paper will go on to address a number of negative refraction examples and will highlight some of the controversy that created such a lot of publicity for this area of metamaterials. To complete the introduction it is necessary to move on nearly half a century from Schuster to 1945. This year sees Mandelshtam setting out the rules obeyed by negative refractive media (Mandelshtam 1945), although he is apparently unaware of the pioneering work of Schuster. One implication of negative refraction is the possibility of creating perfect lenses and much has been made of this in the last few years through some very elegant work (Pendry 2003). Basically, negative refraction permits the use of lenses with plane parallel sides (Silin 1978). In that case monochromatic aberrations can be eliminated. This dramatic consequence has received a lot of attention in recent times by Pendry. This type of lens behaviour is a direct consequence of negative refraction and will not be addressed further in this paper. The aim instead is to establish the basic nature of negative refraction and discuss any conceptual difficulties surrounding it, rather than elaborate particular applications.

## Negative phase aka left-handed materials

Schuster pointed to an absorption band as a possible medium in which to generate backward waves but was concerned that it is region of high absorption. He did not need to invoke a negative relative permittivity nor, indeed, the kind of relative permeability normally associated with purely magnetic materials. The creation of backward waves, however, also flows as an elementary consequence of Maxwell's equations for a lossless, *isotropic* medium. What is required, first of all, is a dispersive relative permittivity $\varepsilon(\omega)$ coupled to a dispersive relative permeability $\mu(\omega)$, where $\omega$ is the angular frequency. The next step is to set $\varepsilon(\omega)$ and $\mu(\omega)$ to be *simultaneously negative* (Veselago 1968) in a certain frequency range to create a transparent band of operation. For example, it has been known for a long time (Izyumov & Medvedev 1966) that electromagnetic waves in ferrites can be modelled by requiring the simultaneous presence of $\varepsilon(\omega)$ *and* $\mu(\omega)$. There may well be frequency ranges in which they are simultaneously negative, giving a degree of transparency, but then there will be the added complication of anisotropy.

As discussed by Veselago, adopting *scalar functions* $\varepsilon(\omega)$ *and* $\mu(\omega)$, simultaneously setting them to be *negative*, and then proceeding with a plane wave solution to Maxwell's equations leads to some elementary conclusions. Assume that plane waves $\exp i(\mathbf{k}\cdot\mathbf{r} - \omega t)$ are propagating, where $\mathbf{k}$ is the wave vector, $\mathbf{r}(x, y, z)$ is the spatial vector involving the coordinates $(x, y, z)$, and $t$ is the time coordinate. Two of Maxwell's equations then yield

$$\mathbf{k} \times \mathbf{E}(\omega) = \omega \mu_0 \mu(\omega) \mathbf{H}(\omega)$$
$$\mathbf{k} \times \mathbf{H}(\omega) = -\omega \varepsilon_0 \varepsilon(\omega) \mathbf{E}(\omega) \quad (3)$$

where $\mathbf{E}(\omega)$ and $\mathbf{H}(\omega)$ are complex and are the Fourier transforms of the field vectors. Hence the Poynting vector $\mathbf{S}$ is

$$\mathbf{S} = \frac{1}{2}\left(\frac{1}{\omega \mu_0 \mu(\omega)}\right)\mathbf{k}|\mathbf{E}|^2 = \frac{1}{2}\left(\frac{1}{\omega \mu_0 \varepsilon(\omega)}\right)\mathbf{k}|\mathbf{H}|^2 \quad (4)$$

The immediate danger here is that this is only a plane wave calculation and it can lead to some confusion, when Snell's law is applied. This point will be put on one side until worries about causality are introduced further on in this article. For the moment, it is apparent that, under the condition $\mu(\omega) < 0$ and $\varepsilon(\omega) < 0$, $\mathbf{k}$ is *anti-parallel* to $\mathbf{S}$, and that *backward waves* can be expected in an isotropic medium for which $\mu(\omega) < 0$ and $\varepsilon(\omega) < 0$. Nothing has been assumed about the refractive index, nor is this strictly necessary since it is what can be called a *derived concept* and does not appear in Maxwell's equations. It can be set equal to the positive root $n = \sqrt{\mu(\omega)\varepsilon(\omega)}$, or the negative root. The choice does not affect the necessary appearance of backward waves but it is merely derived from the fact that any square root offers the options ± (Pokrovsky & Efros 2002). A familiarity with vector analysis leads to a question about the vector products in (3). For $\mu(\omega) > 0$ and $\varepsilon(\omega) > 0$, a *right-handed* rotation about $\mathbf{H}(\omega)$ takes $\mathbf{k}$ to $\mathbf{E}(\omega)$. For $\mu(\omega) < 0$ and $\varepsilon(\omega) < 0$, however, it is a *left-handed* rotation about $\mathbf{H}(\omega)$ that takes $\mathbf{k}$ to $\mathbf{E}(\omega)$. The handedness refers to the rotation of this vector set and *not* any lack of symmetry in the material. In order to possess handedness (Jaggard, Mickelson & Papas 1979) a material must not be capable of being changed by a symmetry operation into its mirror image. Sugar solution is a familiar example of something exhibiting chirality, but then so is a golf club. To say that the left-handed (Veselago 1968) description of backward wave phenomena is a misunderstanding is an understatement but the literature has put this spin upon it and it will be hard to displace. In some ways, it is a similar development to the use of anomalous dispersion, over the last century, when it is plainly not anomalous at all. A much better description is to say it is negative phase medium (Lakhtakia 2003b).

An illustration of the use of negative phase media in refractive situations will now be shown. Each figure is the outcome of a finite-difference time-domain (FDTD) simulation (Taflove & Hagness 2000). First, the electromagnetic equivalent of dropping a stone into a pool is shown in Fig.4. A point source is generated in positive phase free space. The pulse spreads out as a forward wave so the phase waves are travelling in the same direction as the electromagnetic ripples. The pulse eventually interacts with an impedance- matched slab (Ziolkoswki 2003) of negative phase material modelled in the FDTD through a Drude model for the permittivity and permeability with losses set to zero. The spatial grid used is square, the Courant number equal to 0.5 and the source is a hard source located at (126,250). The linear dispersive left-handed, or negative phase, slab is located between x = 250 and 500 and there is no restriction along the y axis. Figure 4a shows that the effect of the negative phase slab is to make the already diffracting beam converge to a single point. This is because of negative refraction.

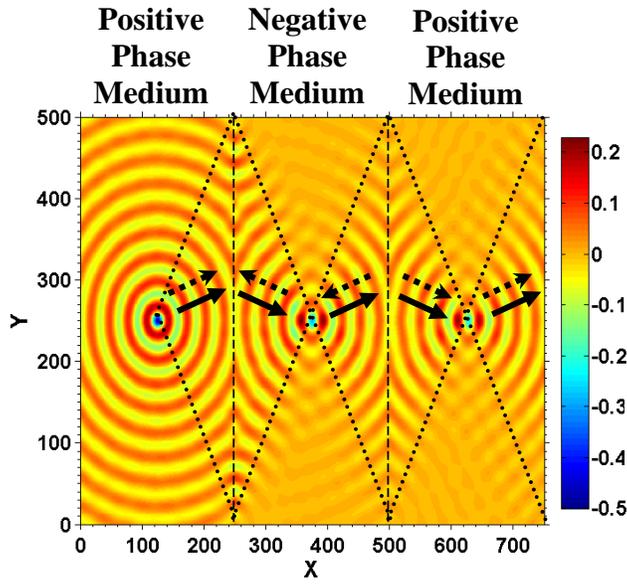

Figure 4a. Point source located in positive phase medium interacting with a slab of negative phase material located between x = 250 and 500 for the whole y-axis, the source is located at the point (126,250). Group and phase velocities indicated by solid or dotted arrows respectively.

It is important to note in any numerical study, in this case the focusing properties of a negative phase slab, that in most cases the simulation is terminated by some form of absorbing boundary, which in essence accelerates the fields to zero, whilst simultaneously minimising reflections. Thus any ray emitted by the aforementioned source that hits the boundary before reaching the interface with the negative phase medium is numerically absorbed and therefore cannot contribute to the quality of the image. In Figure 4a, the regions of refocused energy are indicated by the diagonal dotted lines. Only some of the spectral components of the source are collected by the negative phase medium hence any numerical argument as to the quality of the reformed image must take this into account. Figure 4b demonstrated this fact by doubling the y dimension and essentially repeating the same simulation. In both cases, the simulations were run until the steady-state was reached. It is clear that Figure 4b shows some improvement in the reformed images, even though all the defining physical parameters are unchanged; only the size of the y dimension has been doubled.

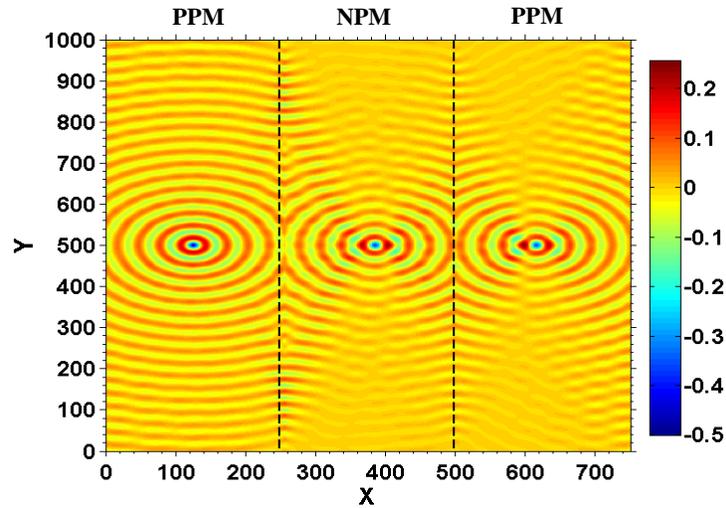

Figure 4b: The effect on image quality of doubling the simulated y dimension. A marginally superior image is formed from the bigger simulation since more of the energy emitted from the source can form an image. Negative phase medium (NPM). Positive phase medium (PPM).

Another interesting example concerns the creation of surface plasmon polaritons (Boardman 1982). Figure 5 shows this type of surface wave being launched using a thin film configuration (Kretschmann 1971) for an incident angle of 60º. The permittivity and permeability of the upper dielectric material are ($\varepsilon$ = 2.3409 $\mu$ = 1), For the negative phase film $\varepsilon$ = -5.8349 $\mu$ = -0.2900. The interfaces of this system are at y = 200 and 220. There is negative refraction involving backward waves.

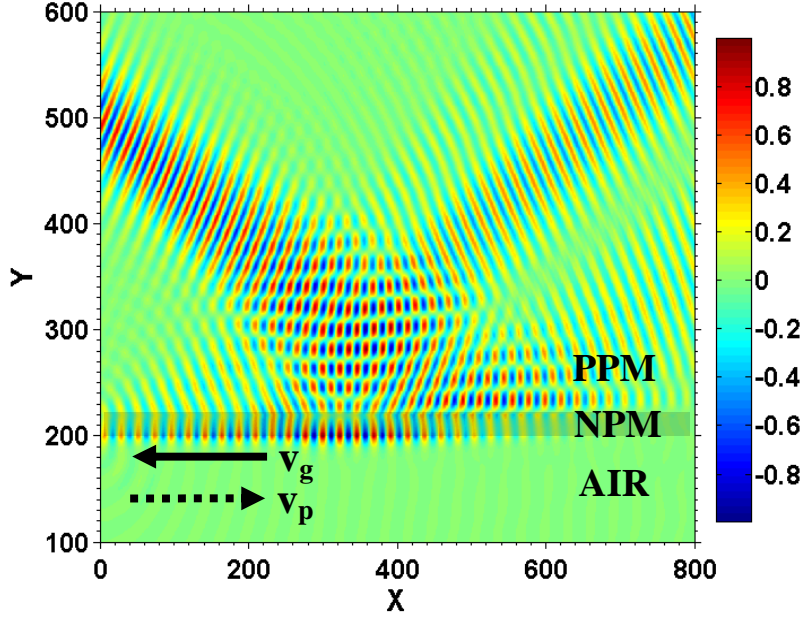

Figure 5: Surface wave formation using the Kretschmann configuration for an incident angle of 60°. PPM ($\varepsilon$ = 2.3409 $\mu$ = 1), NPM ($\varepsilon$ = -5.8349 $\mu$ = -0.2900). Interfaces at y = 200 and 220. Phase velocity is $v_p$ and group velocity is $v_g$.

It must be expected that the underlying condition $\mu(\omega) < 0$ and $\varepsilon(\omega) < 0$ may have to be altered to take into account damping (McCall, Lahthakia & Weiglhofer 2002) that is modelled through the use of, respectively, complex relative permeability and permittivity $\mu = \mu' + i\mu''$, $\varepsilon = \varepsilon' + i\varepsilon''$. To begin with, the square root of the complex number $z = x + iy$ is

$$\sqrt{z} = \pm\sqrt[4]{(x^2+y^2)}\left[\cos\left(\frac{\theta}{2}\right) + i\sin\left(\frac{\theta}{2}\right)\right]; \quad \cos(\theta) = \frac{x}{\sqrt{x^2+y^2}}; \quad \sin(\theta) = \frac{y}{\sqrt{x^2+y^2}} \quad (5)$$

Hence,

$$\sqrt{\varepsilon\mu} = \pm\frac{1}{2\sqrt{(\varepsilon'+|\varepsilon|)(\mu'+|\mu|)}}\left[(\varepsilon'+|\varepsilon|)(\mu'+|\mu|) - \varepsilon''\mu'' + i\left\{\mu''(\varepsilon'+|\varepsilon|) + \varepsilon''(\mu'+|\mu|)\right\}\right] \quad (6)$$

The phase velocity is negative when the real part of $\sqrt{\varepsilon\mu} < 0$ and this takes place whenever

$$(\varepsilon'+|\varepsilon|)(\mu'+|\mu|) < \varepsilon''\mu'' \quad (7)$$

After multiplying both sides of (7) by $(\varepsilon' - |\varepsilon|)(\mu' - |\mu|)$, and noting that the left-hand side then just adds up to $\varepsilon''^2 \mu''^2$ the final condition for the existence of a negative phase velocity is

$$(\varepsilon' - |\varepsilon|)(\mu' - |\mu|) > \varepsilon'' \mu'' \tag{8}$$

Another approach is first of all to note that the complex wave number is

$$k = -\frac{\omega}{c}\frac{1}{\sqrt{2}}\left[\sqrt{\varepsilon'\mu' - \varepsilon''\mu'' + |\varepsilon\mu|} - i(\varepsilon'\mu'' + \varepsilon''\mu')\sqrt{\frac{1}{\varepsilon'\mu' - \varepsilon''\mu'' + |\varepsilon\mu|}}\right] \tag{9}$$

The imaginary part of k must remain positive so that backward waves occur whenever (Weiglhofer & Lakhtakia 2003)

$$\varepsilon'\mu'' + \varepsilon''\mu' < 0 \tag{10}$$

The power flow must be positive for a backward wave, within the convention adopted earlier on and this will occur whenever

$$\varepsilon'|\mu| + \mu'|\varepsilon| < 0 \tag{11}$$

This is yet another equivalent criterion that defines the regime of existence for backward waves in a negative phase medium. These criteria are displayed numerically in Figures 6 and 7 using the models (Ruppin 2000; Pendry et al. 1998; Smith et al. 2000)

$$\varepsilon(\omega) = 1 - \frac{\omega_p^2}{\omega^2 + i\Gamma\omega} \quad \mu(\omega) = 1 - \frac{F\omega^2}{\omega^2 - \omega_0^2 + i\Gamma\omega} \tag{12}$$

and data F = 0.56, $\omega_0 = 0.4\omega_p$, $\Gamma = 0.1\omega_p$.

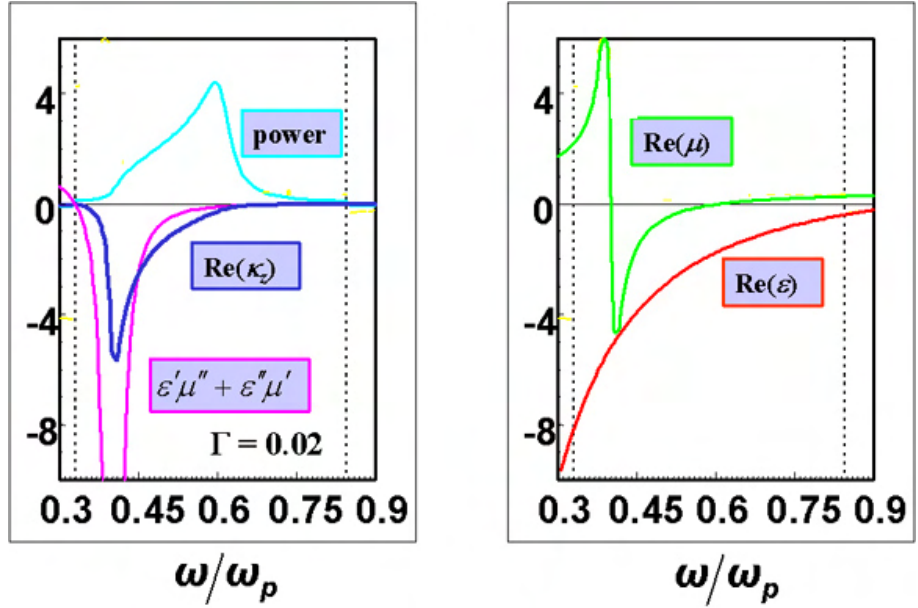

Figure 6: $\Gamma = 0.02$ (a) Plots of the power flow, real part of the wave vector and the condition to have negative refraction (b) Real parts of $\varepsilon$ and $\mu$. Whenever $\varepsilon < 0$ *and* $\mu < 0$ backward waves an occur

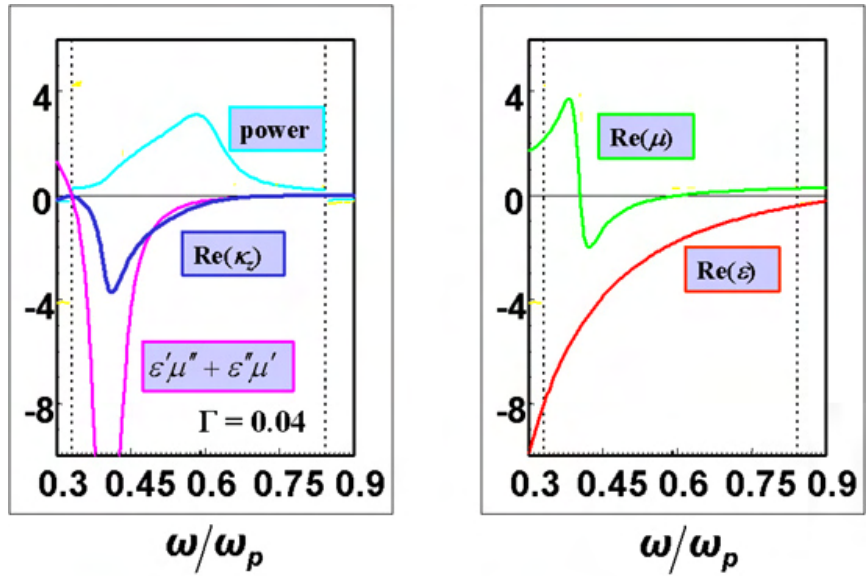

Figure 7: $\Gamma = 0.04$ (a) Plots of the power flow, real part of the wave vector and the condition to have negative refraction (b) Real parts of $\varepsilon$ and $\mu$. When $\mu > 0$ and $\varepsilon < 0$ it is still possible to have backward waves

## Causality

In the preceding discussions it has been tacitly accepted that the negative refraction is perfectly possible in principle and, indeed, observable if only the right kind of experiment can be done, or the right kind of material can be created. That negative refraction should be possible in the absorption bands of a whole host of materials was made clear by Schuster but the expectation of very high losses dampens any enthusiasm for this kind of experiment. If materials with negative relative permittivity and negative relative permeability could exist, or be made as a kind of exotic metamaterial, then backward waves and negative refraction should also be observed. The problem until recently has been to find such materials but the beautiful experiments of Smith and co-workers and the elegant and persuasive papers by Pendry on the possibility of actually creating perfect lenses have provided physical evidence that such materials can be constructed. Yet, not too long ago, a dark cloud of doubt was thrown over the whole area when the very idea of negative refraction, and also the experimental evidence (Smith & Kroll 2000) was questioned (Valanju et al 2002) both on the grounds of interpretation and accuracy. This dark cloud concerns *causality* (Toll 1956) so this, will now be examined in enough detail to reflect the various discussions that have appeared in the literature.

In macroscopic physics the principle that everything has a cause can be safely embraced. Developing the argument further involves the velocity of light and whether it is possible that you can travel backwards in time, or not. Travelling backwards in time has tricky consequences and science fiction writers are fond of using the grandfather paradox (Barjavel 1943), which goes like this. Suppose you travel back in time and prevent your biological grandfather from meeting your grandmother. Then you would have never been conceived, and just as in the film "Back to the Future" you would then have the problem that you would start to fade away through the *lack of a cause*. This paradox can be used to argue that travelling backwards in time must be impossible, if living entities and energy transportation is involved. The fact that no output can be expected before there is an input has been referred to as strict causality (Toll 1956) and can be easily appreciated by imagining that a source of electromagnetic wave is created. Switching on a source, like an aerial, and them pumping the waves out into free space will *cause* electromagnetic waves to propagate out towards an observer, some distance away. The *effect* upon this observer, will be felt at some time later. It is not possible for the observer to access earlier times and, furthermore, the waves cannot travel faster than the velocity of light. Hence, if the observer experiences the waves at any earlier time then the golden rule of Einstein that energy cannot travel faster than the velocity of light will have been violated. It would also imply travelling backwards in time.

When the idea of negative refraction was discussed in modern times it sparked off quite a debate and it all centred upon the cherished principle of causality (Valanju et al. 2002). The relationship of causality to dispersion, is well known and has been thoroughly discussed over the many decades but it is the connection to negative refraction that is so interesting; so this is the focus of attention here. Schuster, who first pointed to the possibility that light incident upon a block of a certain kind of material can be bent by refraction past the normal, was pre-Einstein when his book was published and made no comment, other than to say that it was a curious turn of events. The general principle that there can be materials that can support backward waves is acceptable, but what can be reasonably expected about the refraction process? Also, are the modern objections to negative refraction, apparently aimed at negative phase media, sustainable in any way?

The idea that refraction past the normal is a rational idea can be appreciated by revisiting the whole question but this time using the usual *phase-matching* graphical construction. A flat surface, of infinite extent, separating two semi-infinite isotropic half-spaces will now be considered and in each region the dispersion equation is simply

$$k^2 = \frac{\omega^2}{c^2}\varepsilon\mu \tag{13}$$

where k is the wave number, $\omega$ is the angular frequency, $\varepsilon$ is the relative dielectric permittivity, $\mu$ is the relative magnetic permeability and c is the velocity of light in a vacuum. Equation (13) defines a surface in *k*-space (Saleh & Teich 1991) and Fig. 8 shows that for the isotropic case being considered that the cross-sections are circles.

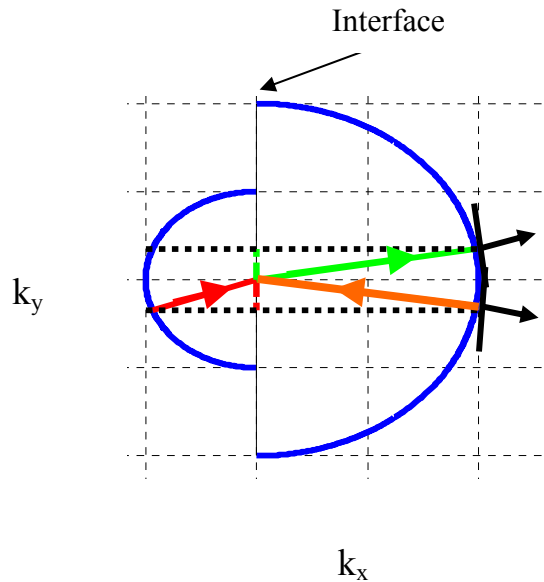

Figure 8 : Wave number surfaces yield circles in the ($k_x$, $k_y$) space selected for this illustration. The interface separates air on the left from a negative phase medium on the right which is sustaining a negative phase velocity. The phase directions are shown and the normals to the tangents surfaces show the energy flow direction.

The wave vectors that are shown represent the *phase* directions. The normals to the surfaces, i.e. the arrows that are perpendicular to the tangents, show the expected direction of the group velocity and hence the energy flow direction. Diagrams like Fig 8 are often used to illustrate phase matching and for a negative phase medium it is the backward flowing phase that must be used to achieve the necessary matching. No phase matching can be achieved by using the forward flowing phase. Once this construction is accepted then the corollary is that the group velocity is anti-parallel to the phase direction and negative refraction is clearly demonstrated. This would appear to the end of the discussion at this point until it realised that this a phase matching diagram, based upon a plane wave and no proper discussion of energy flow has yet taken place. Intuitively, however, it is expected that negative refraction does take place and that the experiments (Smith et al 2000) are beyond reproach. Explicitly, the objections raised a few years ago, to negative refraction, can be set out as follows.

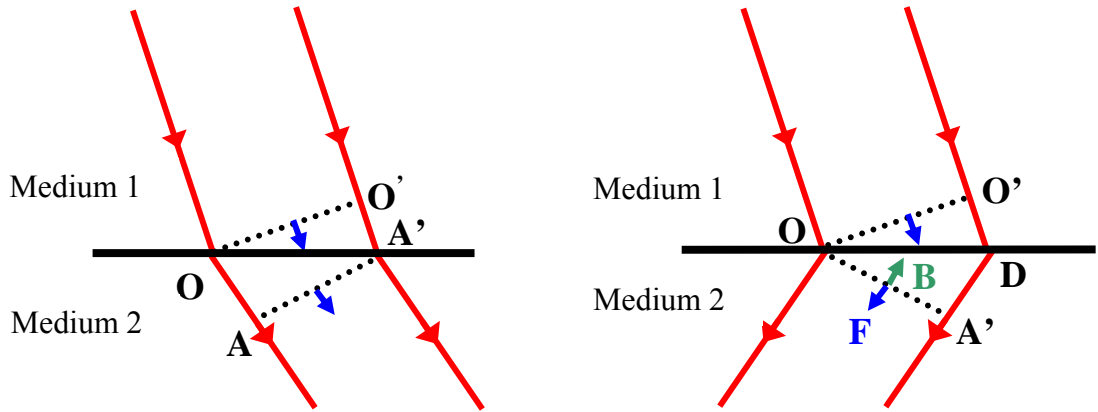

Figure 9 : (a) the positive refraction expected when a beam of light encounters an interface between two positive phase isotropic dielectrics that support forward waves. The wave fronts are dotted lines. The arrows perpendicular to the fronts show the energy flow direction. (b) the same beam of light entering into a negative phase medium. Energy flow in the negative phase medium is in the direction labelled F and the label B denotes a backward phase wave.

Figure 9 adopts the usual method of sketching how refraction works. On the left-hand side two *rays* [carrying energy] defining a beam of light incident upon a flat, unbounded surface are shown. Suppose that the beam approaches the surface through a medium for which $\mu(\omega)$ =1 and $\varepsilon(\omega)$ =constant > 0 and that it then refracts out into air. The wave fronts are perpendicular to the ray directions and point O by the ray on the left. The ray on the right has to make up the distance O'A' but while it is doing this the first ray enters the medium and travels the distance OA. Thus, just at the moment when A' is in position, A is also lined up with it to form the front that is needed for positive refraction. This is entirely expected and there is no reason to suppose that it is unsafe to say that the energy flow is perpendicular to this wave front. The second case on the right is more problematical because the lower half-space consists of a negative phase medium for which $\mu(\omega) < 0$ and $\varepsilon(\omega) < 0$. For the new phase front to be developed in this negative phase medium the outer ray must travel through the distances O'D and DA' in zero time, in order to line up with O, to create the new phase front OA'. To put it bluntly, the point O' must get to A' with infinite speed and this violates the Einstein golden rule that the speed of light cannot be exceeded by an actual energy flow. This apparent violation that the speed of light is finite must also for the reasons given earlier violate causality too i.e. when A' matches up with O to form a front then what is a happening at A' is a cause without an effect. It is like travelling back in time. This is allowed for phase fronts but not for energy rays. The sketch in Figure 9 has been used to claim (Valanju 2002) that negative refraction is not permitted because of the causality violation and the answer to this challenge is quite interesting but it can be countered rather successfully.

On the simplest of levels, the elementary discussion based around a monochromatic plane wave is correct for the phase behaviour because *no energy is carried* by the phase velocity. It is simply enough to say that the negative phase medium carries backward waves that are *moving towards* the surface. This means that Figure 9 may be safely used for phase-matching arguments and the application of Snell's law. The energy flow is a different matter, because Snell's law does not apply to this distribution i.e. there is no phase diagram for the energy flow (Saleh & Teich 1991). Nevertheless, if a separate consideration of the Poynting vector is performed, it ought to be enough to say that the Poynting vector is anti-parallel to the phase vector and that it flows in the direction of the normal to the **k**-space frequency surfaces. To introduce group velocity directly into the description, however, it is necessary to consider more than one plane wave: a beam in other words. A linear beam being used to interrogate a surface is, in fact, an infinite set of plane waves but even a limited set of plane waves may well be the basis of a interesting investigation to see how the energy flows are set up (Pendry & Smith 2002). These authors countered the objection raised to negative refraction by considering the interference pattern created by adding two plane waves, with slightly different frequencies, that are travelling in slightly different directions. This produces a fascinating effect in which the "group" of two waves upon encountering a negative phase medium transmit their energy with a group velocity that is anti-parallel to the phase wave direction but there is also an interference front, more of which will be revealed below.

A reasonable mathematical approach (Maslovski 2002) is to consider the addition of two plane waves under different modulation conditions. In other words, a careful distinction between time-modulation and space- modulation can be made. For the case of time-modulation, consider two waves travelling in the *same direction* that refract through an interface between, for example, air and a negative phase medium. In the latter, two plane waves are also created with slightly different frequencies, $\omega_1$ and $\omega_2$, and slightly different wave numbers, $\mathbf{k_1}$ and $\mathbf{k_2}$. Figure 10 shows the coordinate system and each of these wave vectors in the negative phase medium has the following structure $\mathbf{k} = k_t \hat{\mathbf{x}} + \beta \hat{\mathbf{z}}$, in which $\hat{\mathbf{x}}, \hat{\mathbf{z}}$ are unit vectors. The tangential component of each wave number is $\frac{\omega}{c} \sin(\theta)$ and the normal component, $\beta$, is unknown at this stage but will depend upon the material parameters.

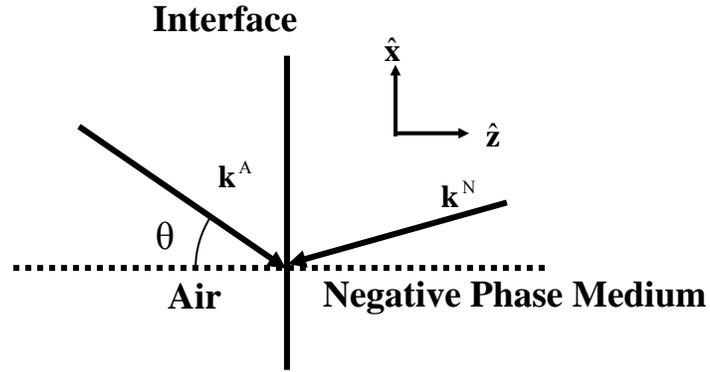

Figure 10: Configuration and coordinate system needed for the time- and space- modulation calculations

If the wave vectors are very close together then $\mathbf{k_2} = \mathbf{k_1} + \delta\mathbf{k}$ and the frequencies are only $\delta\omega$ apart. Using equation (13), and assuming that the negative phase medium is dispersive, then

$$\mathbf{k}_2 - \mathbf{k}_1 \approx \frac{d\mathbf{k}}{d\omega} \delta\omega; \qquad \mathbf{k} \cdot \frac{d\mathbf{k}}{d\omega} = \frac{1}{c^2}\left[\omega\varepsilon\mu + \frac{\omega^2}{2}\frac{d(\varepsilon\mu)}{d\omega}\right] \qquad (14)$$

It is then a simple manipulation to show that

$$\frac{d\mathbf{k}}{d\beta} = \frac{\sin(\theta)}{c}\hat{\mathbf{x}} + \left[\frac{\beta}{\omega} + \frac{\omega^2}{2c^2\beta}\frac{d(\varepsilon\mu)}{d\omega}\right]\hat{\mathbf{z}} \qquad (15)$$

If the maximum of the two-wave combination is considered then expressing the sum as a cosine function shows that the maximum can occur whenever

$$\left(\frac{\omega_2 - \omega_1}{2}\right)t \approx \frac{\delta\omega}{2}t = \left(\frac{\mathbf{k}_2 - \mathbf{k}_1}{2}\right).\mathbf{r} \approx \left(\frac{1}{2}\frac{d\mathbf{k}}{d\omega}.\mathbf{r}\right)\delta\omega \Rightarrow \boxed{t = \frac{d\mathbf{k}}{d\omega}.\mathbf{r}} \qquad (16)$$

The conclusions at this stage are that $\frac{d\mathbf{k}}{d\omega}$ has a tangential component that does not depend upon the material properties and that the expression for t is rather interesting. It is well known that in mathematics that the equation of a plane is $lx + my + nz = p$ where $(l, m, n)$ are the direction cosines of the perpendicular drawn from the origin on the plane and p is the length of the perpendicular. Hence t is the elapsed time and the plane moves along the direction $\frac{d\mathbf{k}}{d\omega}$. In other words normal positive refraction can be expected in this case.

It is now not difficult to imagine the Pendry case in which two waves with the *same carrier frequency* are directed at *slightly different angles* to the surface. This will be called space modulation. Equation (13) gives immediately the condition $\mathbf{k}.\delta\mathbf{k} = 0$. Hence, if the wave vector in the negative phase medium is $\mathbf{k}^N = (k_t, k_z^N)$ and the wave vector in the incident medium (air) is $\mathbf{k}^A = (k_t, k_z^A)$ then

$$\delta\mathbf{k}^N = \left(\hat{\mathbf{x}} - \frac{k_t}{\beta}\hat{\mathbf{z}}\right)\delta k_t \qquad \delta\mathbf{k}^A = \left(\hat{\mathbf{x}} - \frac{k_t}{k_z^A}\hat{\mathbf{z}}\right)\delta k_t \qquad (17)$$

These relationships show that the tangential components are identical and so go smoothly from one to the other as the interface is crossed. The normal components are different and with one of them being controlled by the sign of β and heralds the appearance of negative refraction. A combination of two plane waves with unit amplitude, same frequency, and propagating in slightly different directions has an amplitude $2\cos\left(\frac{\delta\mathbf{k}.\mathbf{r}}{2}\right)$ in which $[\mathbf{k}=\mathbf{k}^A \quad or \quad \mathbf{k}=\mathbf{k}^N]$. This has a maximum whenever the argument of the cosine vanishes. Given the limited information that can be garnered from this mathematical approach it is important to set up a rigorous simulation that embodies both time and space modulation, as shown in figure 11.

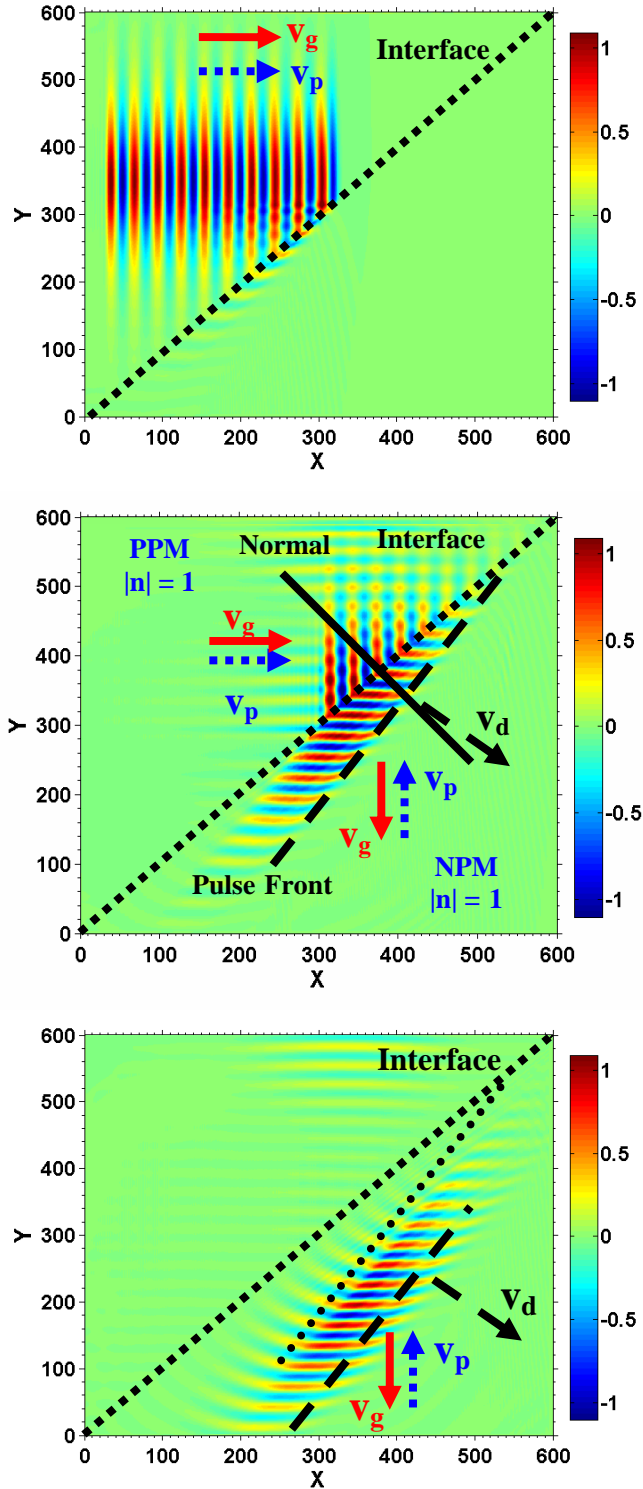

Figure 11: FDTD simulation of a space- time-modulated excitation as it crosses an interface between air and a negative phase medium. For convenience the interface is impedance matched. The figures are snapshots of how the input pulse is evolving for three different times. Note that this type of pulse possesses a large number of frequency components.

It is perhaps worth re-emphasising that these FDTD results are powerful and rely only upon unadulterated Maxwell's equations and record time and space outcomes in a dramatic fashion. Accordingly, Figure 11 reveals (Pendry & Smith 2002; Maslovski 2002) that an incoming pulse changes its shape as it goes through an interface. For greater clarity, Figure 11 opts for an interface between a positive and negative phase medium that is inclined at an angle to the incoming space-time pulse. From the preceding mathematical logic, it is expected that the time-modulation will produce a front that moves in a positive refraction direction and that the space-modulation will produce negative refraction of the actual energy flow. Figure 11 shows precisely this turn of events. The pulse does become distorted. The energy flow, labelled with an arrow showing the direction of the group velocity, is in the negative refraction direction. Finally, the front associated with the time-modulation propagates with a velocity $v_d$ that is set at an angle to the group velocity vector. This displacement velocity causes the distorted pulse to slip sideways so the whole motion has the appearance of a *crab-like* walk (Pendry & Smith 2002). The latter can easily be generated by drawing parallel lines on two separate transparencies and then moving them over each other with the lines inclined at an angle. It will then be observed that there is an interference pattern slipping over the combined pattern formation.

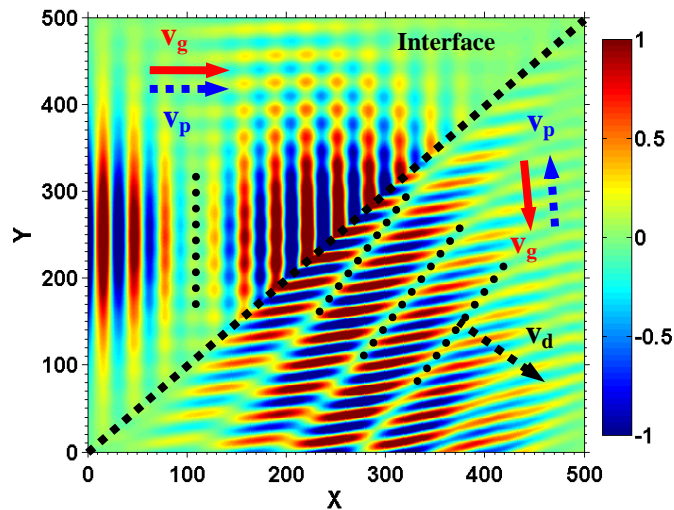

Figure 12: Incident modulated beam in which there are two predominant carrier frequencies. The interface is between air and a negative phase medium. Note the highly visible subsequent *positive* refraction of the temporal interference fringes. The minima have been extenuated by dotted lines.

In Figures 11 and 12 it can be seen that the FDTD method also produces the phase waves and that these are flowing anti-parallel to the group velocity. This is the classic hallmark of backward wave behaviour. All of this discussion points to the fact that the original objection to negative refraction in negative phase media cannot be sustained. The outcome shown in Figures 11 and 12 are very revealing and address any doubts that may be entertained about the behaviour of negative phase media. In fact, the original objection was aimed at negative phase media because writers have sometimes associated negative refraction *entirely* with negative phase. However, anisotropic crystals can support negative refraction. Furthermore, this refraction is associated with *forward* propagating phase waves; using a negative phase medium is not a necessary condition for negative refraction. The current section leads to the acceptance that negative refraction does not violate causality so, having accepted that this is the case, the next section will exploit the familiar wave vector surfaces to explain how anisotropy also leads to negative refraction in a positive phase medium.

### Anisotropic and photonic crystals

Now that the discussion of causality has been brought to a successful conclusion, attention will be turned to crystal anisotropy and to a brief look at photonic crystals. The outcomes for beams or pulses interacting with crystal surfaces needs a full FDTD simulation, if detailed information is required. What the general conclusions show, however, is that a construction based upon rays that carry energy (to which Snell's law *cannot* be applied), and wave fronts that are normal to a wave vector **k** (to which Snell's law *must* be applied) can be used to gain information about the type of refraction taking place. This is an important conclusion because anisotropy effects are a considerable complication beyond isotropy.

The concept of negative refraction is so fresh and exciting in some areas that it may lay claim to being a new frontier in science. Statements surrounding the possible appearance of negative refraction can be exaggerated, however, and sometimes misleading; at best limited in scope. For example, anisotropic crystals, such as calcite, can exhibit amphoteric (from the Greek *amphoteros*, meaning each of two) refraction (Zhang et al. 2003) i.e. refraction can occur positively or negatively. The linkage of this to the general bandwagon associated with negative phase media is incorrect, though, because, as will be seen below, this is a forward wave, or positive phase, phenomenon. Equally, it is wrong to say that an anisotropic crystal cannot be exhibiting negative refraction simply because it is not a negative phase phenomenon (Bliokh, K.Y. & Bliokh, Y.P. 2004). It is perhaps not too surprising then that negative refraction at the interface between two anisotropic crystals has been claimed as being the same kind of negative refraction (Zhang et al. 2003) that the backward wave community is talking about, however misleading this interpretation is. This is because of the general impression created that all negative refraction is connected to backward wave phenomena. On the contrary, this section will emphasise that negative refraction has always been with us in almost every day examples, involving *forward waves* in anisotropic crystals.

To address the issues it is always necessary to emphasise the distinction between rays and wave fronts (Saleh & Teich 1991). As previously stated, rays transport energy whilst fronts point in the direction of the phase velocity. This is illustrated by the familiar example of the kind of double refraction exhibited by a calcite crystal. Double refraction and anisotropy are, of course, inextricably linked and this leads to ordinary and extraordinary rays having orthogonal polarisations. If the dispersion equation for light propagation in a crystal is $\omega = f(\mathbf{k})$, where $\omega$ is the angular frequency and $\mathbf{k}=(k_x, k_y, k_z)$ is a wave vector, then what can be called a $\mathbf{k}$ surface is the surface over which $\omega$ is a constant. This surface is a very useful representation, not least because the direction of the normal to a tangent to this surface is $\nabla_\mathbf{k} \omega(\mathbf{k})$ and that this quantity is $\mathbf{v_g}$, the group velocity. The Poynting vector is proportional to $\mathbf{v_g}$, so the energy rays will point in the direction of the normals to the $\mathbf{k}$ surface, while the phase velocity will be parallel to $\mathbf{k}$. Uniaxial symmetry will be adopted and the $\mathbf{k}$ surfaces, or wave fronts, are spherical for the ordinary wave and non-spherical for the extraordinary wave. Both types have the optic axis as an axis of symmetry, so the inclination of this optic axis to any interface will be of crucial importance to the refraction process. For a uniaxial crystal the $\mathbf{k}$ surface of constant phase for the extraordinary wave is actually elliptic. For the ordinary wave it is spherical. Figs.13a and 13b show, for a given surface the optic axis of the crystal can be inclined and a projection of the **k** space can be drawn using the $(k_x, k_y)$ plane, for example. Matching this projection to air across a flat boundary immediately shows the refraction possibilities.

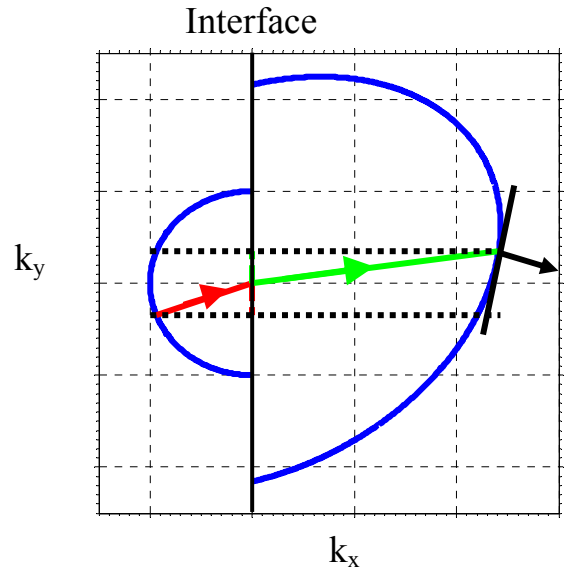

Figure 13 a: Negative refraction at the interface between air and an anisotropic crystal. Note that the normal to the tangent shows the group velocity direction and **k** gives the phase velocity direction.

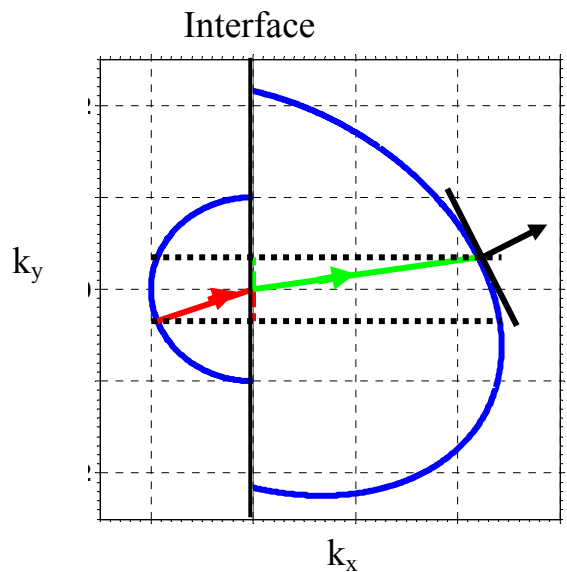

Figure 13b: Positive refraction at the same air/anisotropic interface.

For both the positive and negative refraction the phase waves are travelling forward but at an angle to the group velocity direction. Hence this is a *positive phase* medium, regardless of whether positive or negative refraction is taking place. Negative refraction in a calcite crystal is in fact a familiar property and shows up through the track of the extraordinary ray. It is very easily seen by simply handling a sample of calcite. Negative refraction in positive phase (right-handed) materials has always been with us then. However, note that the frequency surfaces associated with the extraordinary waves in calcite are not very elliptic. This means and that there is only a narrow range of angles of incidence in which negative refraction can take place, with the direction of **k** and the ray velocity being noncollinear. Figure 14 shows a calculation of how the angle of the transmitted energy varies with the optic axis.

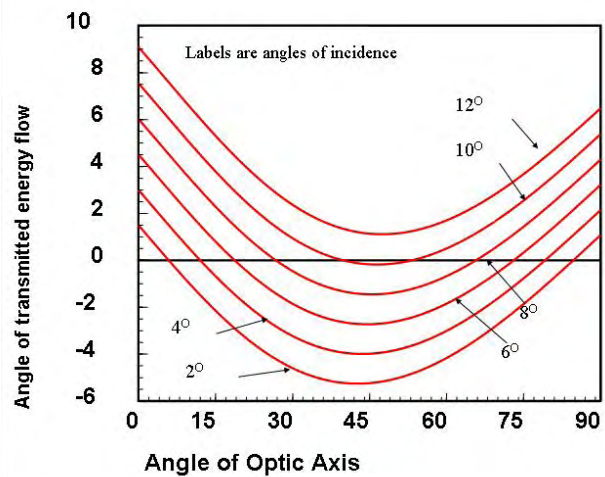

Figure 14 : Calcite can sustain negative refraction.

The small range of negative refraction possessed by calcite makes it more difficult to show up the process using an FDTD approach. This is because such calculations are done in real time over grid spacings that are less than wavelength. Naturally, the range is more than enough to make the track of an extraordinary light beam very visible over a typical crystal length of cm or more. Because of this computational fact, calomel has been selected for the actual simulation to make the negative refraction more visible graphically over rather fewer computational cells than calcite would demand. The results from an FDTD simulation for calomel are displayed in Figures 15 and 16.

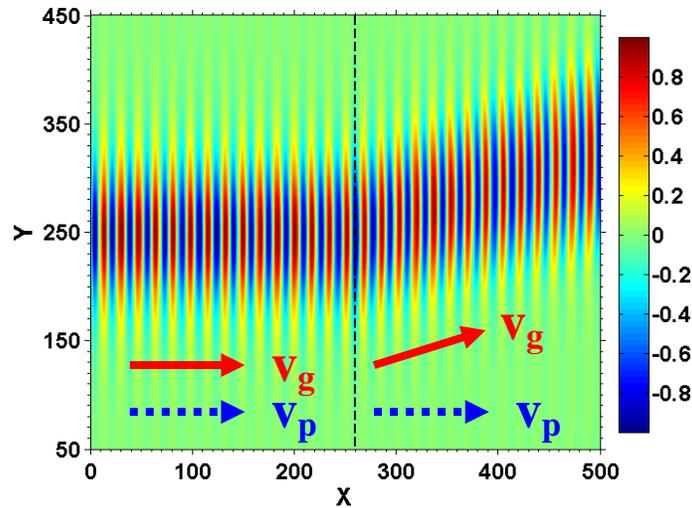

Figure 15: Generation of an extraordinary ray for a calomel crystal with the optic axis inclined at 37º to the y- axis. The beam is incident from air beam is normal incidence from air

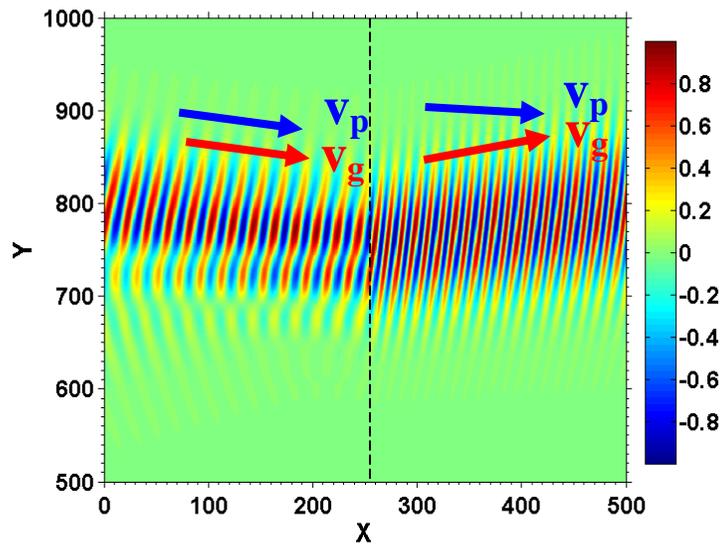

Figure 16 : Extraordinary ray for a calomel crystal slightly inclined from the previous case of normal incidence to demonstrate the occurrence of negative refraction

These figures show that the negative refraction is accompanied by a *forward* travelling phase velocity but this phase velocity is moving at an angle to the group velocity. This positive phase refraction is quite different from the negative phase work that is currently catching such a lot of attention. Just as was discovered from Schuster's work, negative refraction is not a new idea, nor is it very surprising.

Meanwhile the search for isotropic materials with backward wave properties gathers momentum and negative refraction continues to fascinate.

One area of fascination is the extent to which negative refraction can be generated by photonic crystals. Indeed, there is a strong desire for engineered materials exhibiting negative refraction at optical frequencies. These are expected to show more application flexibility than can be offered by natural crystals. Approaches to synthetic optical range negatively refracting media have led to the pursuit of negative phase metamaterials involving nanowires (Podolskiy, Sarychev, Shalaev 2003) but a photonic crystal solution is also very attractive (Foteinopoulou, Economou, & Soukoulis, 2003; Luo et al. 2002). Although photonic crystals are very interesting only their bearing upon the negative refraction perspective will be addressed here. It has been established that some features of photonic crystals produce negative refraction similar to that occurring in isotropic negative phase metamaterials. However, it appears that a degree of caution is needed because there is more than one mechanism for negative refraction in photonic crystals. This is rather different from the properties outlined above for the isotropic negative phase materials that lead directly to antiparallel group and phase velocities. In photonic crystals, negative refraction can occur even though it is still effectively a positive phase medium. Negative refraction can occur as a result of anisotropy, in a similar manner to the birefringent crystals discussed earlier. In other words, the gradient of an equifrequency contour in a photonic crystal need not be parallel to the wave vector, leading to forward wave negative refraction. In particular, for this type of photonic crystal operation, negative refraction will only occur under correct matching of the wave vectors and may not be allowed for certain directions through the crystal (Gralak, Enoch & Tayeb 2000). Also, positive phase negative refraction can occur due to coupling to a higher order Bragg wave (Foteinopoulou, Economou, & Soukoulis 2003). In spite of these concerns, it should be expected that the constant-frequency dispersion curves are circular around the centre of the first Brillouin zone. In that case, the usual negative phase behaviour may manifest itself if the frequency gradient is suitable. Even if this achieved, there is often residual anisotropy that can cause the Poynting vector and the wave vector to be non-collinear. An activity made popular by Pendry is 'superlensing'. As pointed out earlier, when briefly referring to a 'superlens' as a lens with plane parallel sides (Silin 1978), monochromatic aberrations can be eliminated through its use. There is a fundamental refocusing of a source by a slab of negatively refracting material. It would be very nice if this could be done with a photonic crystal and basically it can but there is some disagreement as to the mechanism involved. Some researchers attribute the refocusing to negative refraction, but perhaps it is due to a tunnelling effect and the explanation is much more complicated.

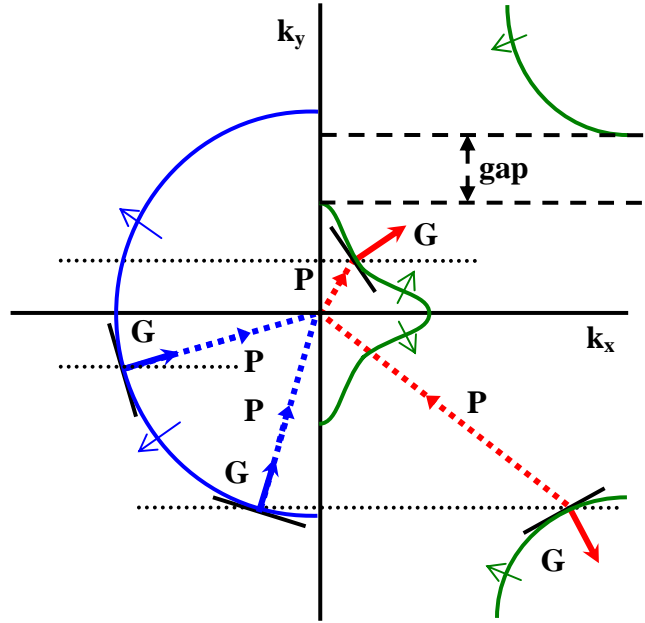

Figure 17: Diagram of wave number surfaces and group (G) and phase (P) wave directions for a possible photonic crystal interfaced to an isotropic positive phase medium. The directions of increasing gradient for the equifrequency contours are marked by unlabelled arrows. There is residual anisotropy causing non-parallelism.

This diagram shows two waves incident at different angles from a generic isotropic positive phase medium media onto a possible photonic crystal. In this case, the positive phase medium and the inner contour of the photonic crystal both have a positive gradient, i.e. for increasing frequency these contours will sweep out towards increasing *k* values, and so the phase velocity is *forward* relative to the group velocity. However, the outer contour of the photonic crystal has a negative gradient and hence the phase velocity is *backward* relative to the group velocity. As well as indicating the anisotropic behaviour of photonic crystals, this diagram also demonstrates that, as the angle of incidence increased a gap (a region of no propagation) is encountered. Figure 17 looks complicated but it does in reality imply that negative refraction for this type of photonic crystal is a possibility as clarified below.

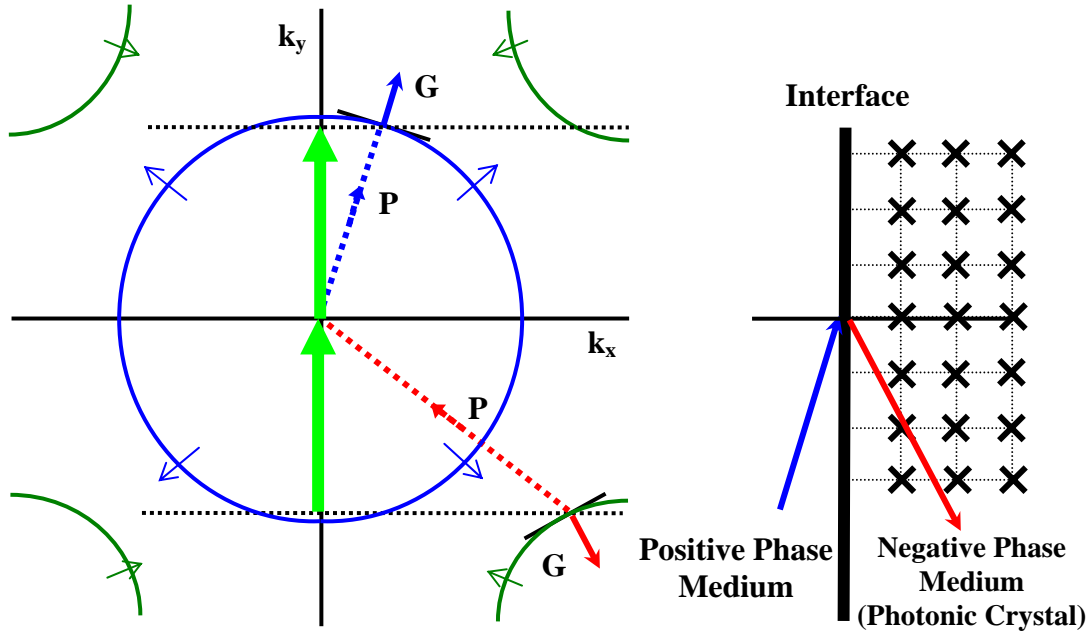

Figure 18: Extended phase matching construction to match isotropic medium to phase behaviour in a stylised square lattice photonic crystal. ✗ denotes a lattice site.

Figure 18 shows the usual phase matching construction mapped onto the *k*-space expected from a simple square lattice selected to represent a typical photonic crystal. Only the outer frequency contours are used in this case because not only are they circular but they have the correct frequency gradient to guarantee the generation of negative refraction associated with negative phase velocity (Gralak, Enoch & Tayeb 2000). Note that absolute antiparallelism can not be guaranteed and this is evidence of the residual anisotropy mentioned earlier in the text.

**Conclusions**

This paper provides a perspective on the concept of negative refraction. It is not intended as a comprehensive catalogue of all the papers published in this area. Its purpose is to address the concept of negative refraction and its historical and current background. Beginning with the ideas of Schuster, it is shown that the concept of negative refraction has been in the literature for a very long time but, nevertheless, in many areas of science, the excitement that surrounds this phenomenon is quite recent. It is emphasized that the modern search for negative refraction mainly involves metamaterials but it also can be found in a limited way in anisotropic crystals and, in quite a promising way, in photonic crystals. The latter could also be regarded as metamaterials. The concept of forward and backward waves is clearly illustrated using FDTD simulations in which the direction of the phase front and the energy rays are defined through the nature of the method. It is demonstrated, visually that negative refraction with forward waves occurs in anisotropic crystals and that isotropic media can support backward wave negative refraction. It is hoped that this paper provides the perspective implied by the title.

**Acknowledgements**

The authors would like to thank Dr. Peter Egan for his advice and specific assistance with figures 3, 6, 7 and 14.